\documentclass[pra,showpacs]{revtex4-1}
\usepackage{graphicx}
\usepackage{dcolumn}
\usepackage{bm}
\usepackage{bm}
\usepackage{amsmath}
\usepackage[percent]{overpic}
\usepackage{subfig}
\usepackage{amssymb}
\begin{document}

\title{Remote preparation of $W$ states from imperfect bipartite sources}

\author{M. G. M. Moreno}
\author{M\'arcio M. Cunha}
\author{Fernando Parisio}
\email{parisio@df.ufpe.br}
\affiliation{Departamento de F\'{\i}sica, Universidade Federal de Pernambuco, 50670-901,Recife, Pernambuco, Brazil}


\begin{abstract}
Several proposals to produce {\it tripartite} $W$-type entanglement are probabilistic even if no imperfections are considered in the processes.
We provide a deterministic way to remotely create $W$ states out of an EPR source. The proposal is made viable through measurements (which can be demolitive) in an appropriate three-qubit basis. The protocol becomes probabilistic only when source flaws are considered. It turns out that, even in this situation, it is robust against imperfections in two senses: (i) It is possible, after postselection, to create a pure ensemble of $W$ states out of an EPR source containing a systematic error; (ii) If no postselection is done, the resulting mixed state has a fidelity, with respect to a pure $|W\rangle$, which is higher than that of the imperfect source in comparison to an ideal EPR source. This simultaneously amounts to entanglement concentration and {\it lifting}.
\end{abstract}

\maketitle

\section{\label{sec:level1}Introduction}
The inventory of potential achievements that quantum entanglement may bring about has steadily grown for decades. It is the concept behind most of the non-trivial, classically prohibitive tasks in information science \cite{nilsen}. 
However, for entanglement to become a useful resource in general, much work is yet to be done. The efficient creation of entanglement, often involving many degrees of freedom, is one of the first challenges to be coped with, whose simplest instance are the sources of correlated pairs of two-level systems. These sources have been used to demonstrate the possibility of teleporting an unknown qubit \cite{teleport} and to disclose the nonlocality of quantum mechanics \cite{nonlocal}. Some other important tasks require more involved kinds of entanglement, e. g., a prominent framework for measurement-based quantum computation is possible only if cluster sates are available \cite{Raussendorf}. 

In the last two decades the increasing interest in complex entanglement motivated a fair amount of works aiming at the creation of larger than Einstein-Podolsky-Rosen (EPR) states. Most of these efforts have produced Greenberger-Horne-Zeilinger (GHZ) states \cite{ghz} involving three \cite{ghz3}, four \cite{ghz4}, five \cite{ghz5}, six \cite{ghz6}, and eight \cite{ghz8} qubits. Other multiqubit states have been built, though more sparsely, for instance, Dicke states \cite{dicke6} and graph states \cite{ghz6}. Intermediate kinds of entanglement have been reported in Ref. \cite{xin}. 

In this manuscript we are mainly interested in the creation of tripartite $W$ states. 
In the first experimental realization of such a state \cite{Manfred}, four photons originated from second order parametric down conversion (PDC) are sent to distinct spatial modes and through linear optical elements. The conditional detection of one of the photons leaves the remaining three photons in the desired $W$ state with a probability of $1/32$ for each second order PDC event. Since these occurrences are by themselves rare, the whole process lacks efficiency.
A sufficiently robust experiment to enable state tomography is described in Ref. \cite{mikami2}, where three-fold coincidences were observed with a rate about 40 times higher than that of \cite{Manfred}. However, the creation of $W$ states remained probabilistic, thus, requiring postselection. In Ref. \cite{mikami}, an experiment based on two independent first order PDC events is described, resulting in $W$ states of four and three photons. The latter being achieved, again, only probabilistically,  through a measurement on one of the parties. Projective measurements, that present intrinsic stochasticity, can also be used to produce $W$-type entanglement \cite{wieczorek}. More in the spirit of the present work, bipartite entanglement can be considered as an available resource, as is the case of Ref. \cite{tashima}, that takes two pairs of photons in EPR states as the building blocks to probabilistically produce $W$ states of three parties.
Very recently general tripartite entangled states were encoded in the nuclei of the fluorine atoms of trifluoroiodoethylene molecules, via nuclear magnetic resonance \cite{nmr}. This technique, however, is unsuited for remote preparation because the state is encoded in a spatially localized structure. 

In the theoretical front, protocols based on PDC \cite{Yamamoto}, linear optics \cite{cheng,tashima1,tashima2,tashima3}, sources of EPR states \cite{xu}, atomic systems \cite{yang}, and  nitrogen-vacancy centers \cite{tong} can be found, all being probabilistic. 
In contrast, the authors of Ref. \cite{ozaydin} propose a deterministic scheme to create four-partite $W$ states, which can be extended to generate $W$ states of arbitrary dimension \cite{yesil}.
More recently, a procedure was suggested to create entangled states of several degrees of freedom, which is particularly suited for $W$ states. This fusion operation \cite{fusion0} corresponds to a swapping procedure and employs two entangled states with dimension $d$ to produce an entangled state of dimension $D=2d-2$. 
\begin{figure}
\includegraphics[width=7.5cm]{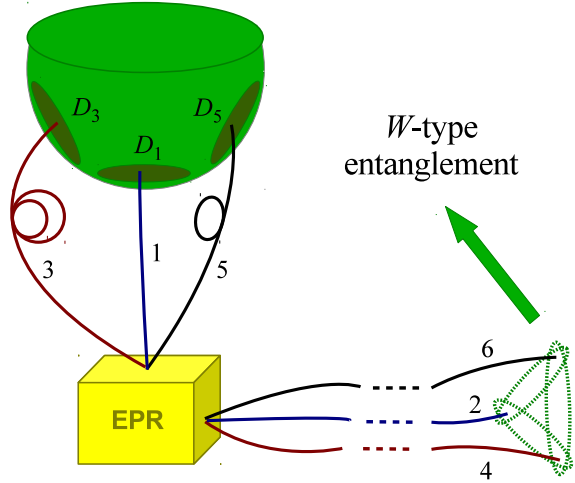}
\caption{(color online) EPR entangled pairs (1-2, 3-4, and 5-6) produced in sequence in a single source are spatially separated. Odd-labeled particles are appropriately measured, after which, the remaining particles are left in a genuine tripartite entangled state.}
\label{fig1}
\end{figure}
\section{Deterministic production of $W$ states}
We assume that there is a source of bipartite entanglement, ideally delivering identical EPR pairs. Three such doubles are needed in each run, with one particle of each pair being sent to three detectors ($D_1$, $D_2$ e $D_3$). 
At this point let us simply consider that the state of three pairs coming from the source is $| \phi^+ \rangle^{\otimes 3}$:
\begin{equation}
\label{source}
\frac{1}{2\sqrt{2}}(|00\rangle_{12}+|11\rangle_{12})\otimes(|00\rangle_{34}+|11\rangle_{34})\otimes(|00\rangle_{56}+|11\rangle_{56})\;. 
\end{equation}
Of course, it does not matter which ERP state one choses, provided that it is known.
We intend to make a triple measurement on the odd labelled particles (subsystem $A$) which are sent to the detectors, see Fig. \ref{fig1}. 
We note from the outset that the measurement on subsystem $A$ may be destructive and will be assumed to be so, although this is not a logical necessity. The key point of our protocol, which enables the deterministic creation of $W$ states, is the the following set of kets:
\begin{eqnarray}
\label{Wbasis}
\nonumber
|W_1\rangle&=&\frac{1}{\sqrt{3}}(|001\rangle+|010\rangle+|100\rangle)\,,\\
\nonumber
|W_2\rangle&=&\frac{1}{\sqrt{3}}(|000\rangle-|011\rangle+|101\rangle)\,,\\
\nonumber
|W_3\rangle&=&\frac{1}{\sqrt{3}}(|000\rangle+|011\rangle-|110\rangle)\,,\\
\nonumber
|W_4\rangle&=&\frac{1}{\sqrt{3}}(|000\rangle-|101\rangle+|110\rangle)\,,\\
\nonumber
|W_5\rangle&=&\frac{1}{\sqrt{3}}(|110\rangle+|101\rangle+|011\rangle)\,,\\
\nonumber
|W_6\rangle&=&\frac{1}{\sqrt{3}}(|111\rangle-|100\rangle+|010\rangle)\,,\\
\nonumber
|W_7\rangle&=&\frac{1}{\sqrt{3}}(|111\rangle+|100\rangle-|001\rangle)\,,\\
|W_8\rangle&=&\frac{1}{\sqrt{3}}(|111\rangle-|010\rangle+|001\rangle)\,.
\end{eqnarray}
The state $|W_1\rangle (\equiv |W\rangle)$ is the tripartite $W$ state as we are used to it and $|W_5\rangle$ its negation. The other six states may not seem to be of the same category, since they are not combinations of kets with the same number of ``excitations''. However, it is easy to see that they also represent perfect $W$ states, for, $|W_1\rangle$ is obtainable from any of them via local, unitary operations, e. g., $|W_1\rangle=(\mathbb{I}\otimes\sigma_{z}\otimes\sigma_{x})|W_2\rangle=(\sigma_{z}\otimes\sigma_{x}\otimes\mathbb{I})|W_3\rangle$, etc. Note also that $|W_{j+4}\rangle=\hat{\sigma}_x^{\otimes3}|W_j\rangle$ and $\langle W_j|W_k\rangle=\delta_{jk}$. Therefore, we have at our disposal an orthonormal $W$-state basis, that, in principle, corresponds to the eigenvectors of some observable that can be measured. 
Given the ideal source state $| \phi^+ \rangle^{\otimes 3}$, after conveniently reordering the kets, it can be written as
\begin{eqnarray}
\label{psi}
\nonumber
\frac{1}{2\sqrt{2}}[|000\rangle_{A}|000\rangle_{B} + |001\rangle_{A}|001\rangle_{B} +|010\rangle_{A}|010\rangle_{B}\\
\nonumber
+ |011\rangle_{A}|011\rangle_{B} +|100\rangle_{A}|100\rangle_{B} +|101\rangle_{A}|101\rangle_{B} \\
\nonumber
+|110\rangle_{A}|110\rangle_{B} + |111\rangle_{A}|111\rangle_{B}]\;,
\end{eqnarray}
where particles 1, 3, and 5 belong to subsystem $A$ and the remaining particles to the subsystem $B$. This state is given in terms of basis (\ref{Wbasis}) in the totally correlated form:
\begin{equation}
\frac{1}{2\sqrt{2}}\sum_{j=1}^8|W_j\rangle_{A}\otimes|W_j\rangle_{B}\,.
\end{equation}
By executing a projective measurement in this basis for particles in $A$ we {\it deterministically} obtain, after local operations and classical communication (LOCC), a perfect set of pure states in the standard form $|W_1\rangle$ in subsystem $B$. Of course, the measurements in the entangled basis (\ref{Wbasis}) constitute a technical difficulty in practice, but there is nothing that prevents their realization, in principle (more on this point in the next sections). The required local unitary operations are shown in table \ref{table1}. Note, in addition, that it does not matter how far apart are the particles of system $B$, thus, the preparation may be remote. 
\begin{table}[h!]
\centering
\begin{tabular}{| c | c |}
\hline
Result in $A$& Operation to be done on $B$\\
\hline
$|W_1\rangle$ & $\hat{O}_1=\mathbb{I}\otimes\mathbb{I}\otimes\mathbb{I}$\\
\hline
$|W_2\rangle$ & $\hat{O}_2=\mathbb{I}\otimes\sigma_{z}\otimes\sigma_{x}$\\
\hline
$|W_3\rangle$ & $\hat{O}_3=\sigma_{z}\otimes\sigma_{x}\otimes\mathbb{I}$\\
\hline
$|W_4\rangle$ & $\hat{O}_4=\sigma_x\otimes\mathbb{I}\otimes\sigma_z$\\
\hline
$|W_5\rangle$ & $\hat{O}_5=\sigma_x\otimes\sigma_x\otimes\sigma_x$\\
\hline
$|W_6\rangle$ & $\hat{O}_6=\sigma_x\otimes i\sigma_y\otimes\mathbb{I}$\\
\hline
$|W_7\rangle$ & $\hat{O}_7=i \sigma_y\otimes\mathbb{I}\otimes\sigma_x$\\
\hline
$|W_8\rangle$ & $\hat{O}_8=\mathbb{I}\otimes\sigma_x\otimes i \sigma_y$\\
\hline
\end{tabular}
\caption{Operations to be done on system $B$ after the measurements on $A$ to deterministically produce $|W_1\rangle$.}
\label{table1}
\end{table}

A more evident, though analogous procedure is possible for GHZ states. The corresponding measurement involves the basis
$|{\rm GHZ}^{\pm}_{ijk}\rangle=(|ijk\rangle\pm|\overline{ijk}\rangle)/\sqrt{2}$, where the overbar denotes negation and $i,j,k=0,1$. 
For the ideal source of states (\ref{source}) one gets the result 
\begin{equation}
\sim |{\rm GHZ}^{+}_{000}\rangle\otimes|{\rm GHZ}^{+}_{000}\rangle+ \dots + |{\rm GHZ}^{-}_{111}\rangle\otimes|{\rm GHZ}^{-}_{111}\rangle\;,
\nonumber
\end{equation}
also enabling deterministic creation of this kind of tripartite state. However, when the source is imperfect, the process for the two classes of states 
lead to quite distinct results, as we discuss below.
\section{Entanglement concentration and lifting}
We are now in position to address a more realistic entanglement source containing a systematic error. The consideration of source flaws is relevant in the promotion of any theoretical proposal into a feasible process. This has been considered, for instance, in an experiment for secure quantum key distribution \cite{feihu} (see also \cite{mizutani, nagamatsu}) and also in state preparation \cite{pulse-mode,scherer}.
The source is now described by
$|\Phi_0\rangle=\left(a |00\rangle_{12} + b|11\rangle_{12}\right)\otimes\left(a |00\rangle_{34} +b |11\rangle_{34}\right)\otimes\left(a |00\rangle_{56} + b|11\rangle_{56}\right)$,
where, hereafter, $a$ is assumed to be real and positive, and $a\ne |b|$. This corresponds to
\begin{eqnarray}
\label{psi1}
\nonumber
|\Phi_0\rangle&=& a^3 |000\rangle_{A}|000\rangle_{B} + a^2b|001\rangle_{A}|001\rangle_{B} +\\
\nonumber
&& a^2b|010\rangle_{A}|010\rangle_{B} + ab^2|011\rangle_{A}|011\rangle_{B} +\\
\nonumber
&& a^2b|100\rangle_{A}|100\rangle_{B} + ab^2|101\rangle_{A}|101\rangle_{B} +\\
&& ab^2|110\rangle_{A}|110\rangle_{B} + b^3 |111\rangle_{A}|111\rangle_{B}\;,
\end{eqnarray}
where we reordered the state, which, in this case, is expressed in terms of basis (\ref{Wbasis}) as
\begin{eqnarray}
\label{Wimperf}
\nonumber
|\Phi_0\rangle= a^2b |W_1\rangle \otimes |W_1\rangle+ab^2 |W_5\rangle \otimes |W_5\rangle\\
\nonumber
+\sum_{k=2,3,4}|W_k\rangle\otimes\left[ab^2|W_k\rangle+\frac{1}{\sqrt{3}}(a^3-ab^2)|000\rangle \right]\\
+\sum_{k=6,7,8}|W_k\rangle\otimes\left[a^2b|W_k\rangle+\frac{1}{\sqrt{3}}(b^3-a^2b)|111\rangle \right]\;.
\end{eqnarray} 
\begin{table*}
\centering
\begin{tabular}{| c | c |c|}
\hline
Result in $B$ after measurement in $A$ & State left in $B$ after LOCC (Table \ref{table1}) & Probability \\
\hline
$|W_1\rangle$ & $|W_1\rangle$& $a^4|b|^2$\\
\hline
$\sim (a^2+2b^2)|W_2\rangle+(a^2-b^2)(|W_3\rangle+|W_4\rangle)$ & $\sim (a^2+2b^2)|W_1\rangle+(a^2-b^2)(|W_8\rangle-|W_7\rangle)$&$a^2(a^4+2|b|^4)/3$\\
\hline
$\sim (a^2+2b^2)|W_3\rangle+(a^2-b^2)(|W_2\rangle+|W_4\rangle)$ & $\sim (a^2+2b^2)|W_1\rangle+(a^2-b^2)(-|W_8\rangle+|W_6\rangle)$ &$a^2(a^4+2|b|^4)/3$\\
\hline
$\sim (a^2+2b^2)|W_4\rangle+(a^2-b^2)(|W_2\rangle+|W_3\rangle)$ & $\sim (a^2+2b^2)|W_1\rangle+(a^2-b^2)(|W_7\rangle-|W_6\rangle)$ &$a^2(a^4+2|b|^4)/3$\\
\hline
$|W_5\rangle$ & $|W_1\rangle$ & $a^2|b|^4$\\
\hline
$\sim (b^2+2a^2)|W_6\rangle+(b^2-a^2)(|W_7\rangle+|W_8\rangle)$ & $\sim (b^2+2a^2)|W_1\rangle+(b^2-a^2)(|W_8\rangle-|W_7\rangle)$ & $|b|^2(|b|^4+2a^4)/3$\\
\hline
$\sim (b^2+2a^2)|W_7\rangle+(b^2-a^2)(|W_6\rangle+|W_8\rangle)$ & $\sim (b^2+2a^2)|W_1\rangle+(b^2-a^2)(-|W_8\rangle+|W_6\rangle)$ & $|b|^2(|b|^4+2a^4)/3$\\
\hline
$\sim (b^2+2a^2)|W_8\rangle+(b^2-a^2)(|W_6\rangle+|W_7\rangle)$ & $\sim (b^2+2a^2)|W_1\rangle+(b^2-a^2)(|W_7\rangle-|W_6\rangle)$ &$|b|^2(|b|^4+2a^4)/3$\\
\hline
\end{tabular}
\caption{States left in $B$ after the measurement in $A$ (first column), and after the LOCC prescribed in Table \ref{table1} (second column). Note that the states $|W_2\rangle$, $|W_3\rangle$, $|W_4\rangle$, and $|W_5\rangle$ do not appear after the unitary operations. The third column shows the probabilities for each outcome. For the sake of clarity the states are not normalized.}
\label{table2}
\end{table*}
Therefore, it remains possible to get an ideal set of $W$ states after postselection, because whenever the result of the measurement is $|W_1\rangle$ ($|W_5\rangle$) the produced state on $B$ is $|W_1\rangle$ ($|W_5\rangle$) {\it no matter} the values of $a$ and $b$. This occurs with probability $|a^2b|^2$ ($|ab^2|^2$). Thus, by postselecting the outcomes $|W_1\rangle$ and $|W_5\rangle$, proceeding the appropriate LOCC for the latter, one gets a pure ensemble of standard $W$ states. The success probability is $P=|a^2b|^2+|ab^2|^2=a^2-a^4$.
Note that it is bounded from above by $P=1/4$, since we are eliminating the other six outcomes even if they lead to states with high fidelity. 

In a more realistic scenario, one may not be able to carry out a full measurement in the basis (\ref{Wbasis}). This is indeed the case of our present technical development. Note however that if we can unambiguously distinguish $|W_1\rangle$ from $|W_5\rangle$, and these from the other states, then we are able to remotely produce perfect $W$ states from an imperfect bipartite source with probability $P=a^2-a^4$. It is possible to tackle an equivalent task with linear-optics elements in the case of a GHZ basis \cite{wei}.

Although we cannot directly compare the entanglement of systems with two and three parties, we argue that the previous procedure leads to  entanglement concentration \cite{conc}. The source of bipartite states may have arbitrarily low entanglement and the result is always a highly entangled tripartite state, of course at the cost of a proportional reduction in the number of elements in the final ensemble. But this is exactly what happens for entanglement concentration between ensembles of states with the same dimensionality \cite{bose}. If the source is composed by $N$ quasi-separable pairs with state $|\psi_{\epsilon}\rangle=\epsilon|00\rangle+\sqrt{1-\epsilon^2}|11\rangle$, then we get $\epsilon^2N/2+O(\epsilon^4)$ perfect $W$ states, asymptotically. It is evident that any reasonable measure of multipartite entanglement $E$, would give $E(|W\rangle)>E(|\psi_{\epsilon}\rangle^{\otimes3})$, for sufficiently small $\epsilon$.  Also, because the entanglement delivered by the source is bipartite while the product presents genuine tripartite entanglement, we say that the entanglement has been {\it lifted} (from a ``small'' Hilbert space to a larger one). This is a remarkable property of $W$ states, and an analogous situation does not exist for GHZ states. If the source has $a \ne |b|$, by any small extent, then the probability to get an exact GHZ state in any run is zero, and no postselection could help in getting an ideal GHZ ensemble. This adds to the reasoning that $W$ states are less entangled than GHZ states, in the sense that from three non-maximally entangled states one can, via stochastic LOCC (SLOCC) \cite{dur} get a perfect $W$ but not a perfect GHZ. 
\section{High-quality $W$ states without postselection}

The referred postselection may be a waste of resources if some amount of error can be tolerated in a scenario where full measurements can be carried out. For a good, but imperfect source, the outputs are either perfect or so close to the ideal states that the chance that they can be distinguished in practice is very small. Note from the second line of Eq. (\ref{Wimperf}), that if, e. g., we obtain $|W_2\rangle$ in subsystem $A$, then the state in $B$ is proportional to $\sqrt{3}|W_2\rangle+(a^2/b^2-1)|000\rangle$. So, for $a^2=1/2+\epsilon$ ($|b|^2=1/2-\epsilon$), with $\epsilon<<1$, we get $\sqrt{3}|W_2\rangle+4\epsilon e^{-2i \theta}|000\rangle + O(\epsilon^2)$, where $b=|b|e^{i \theta}$. 

\begin{figure}[t]
\includegraphics[width=4.25cm]{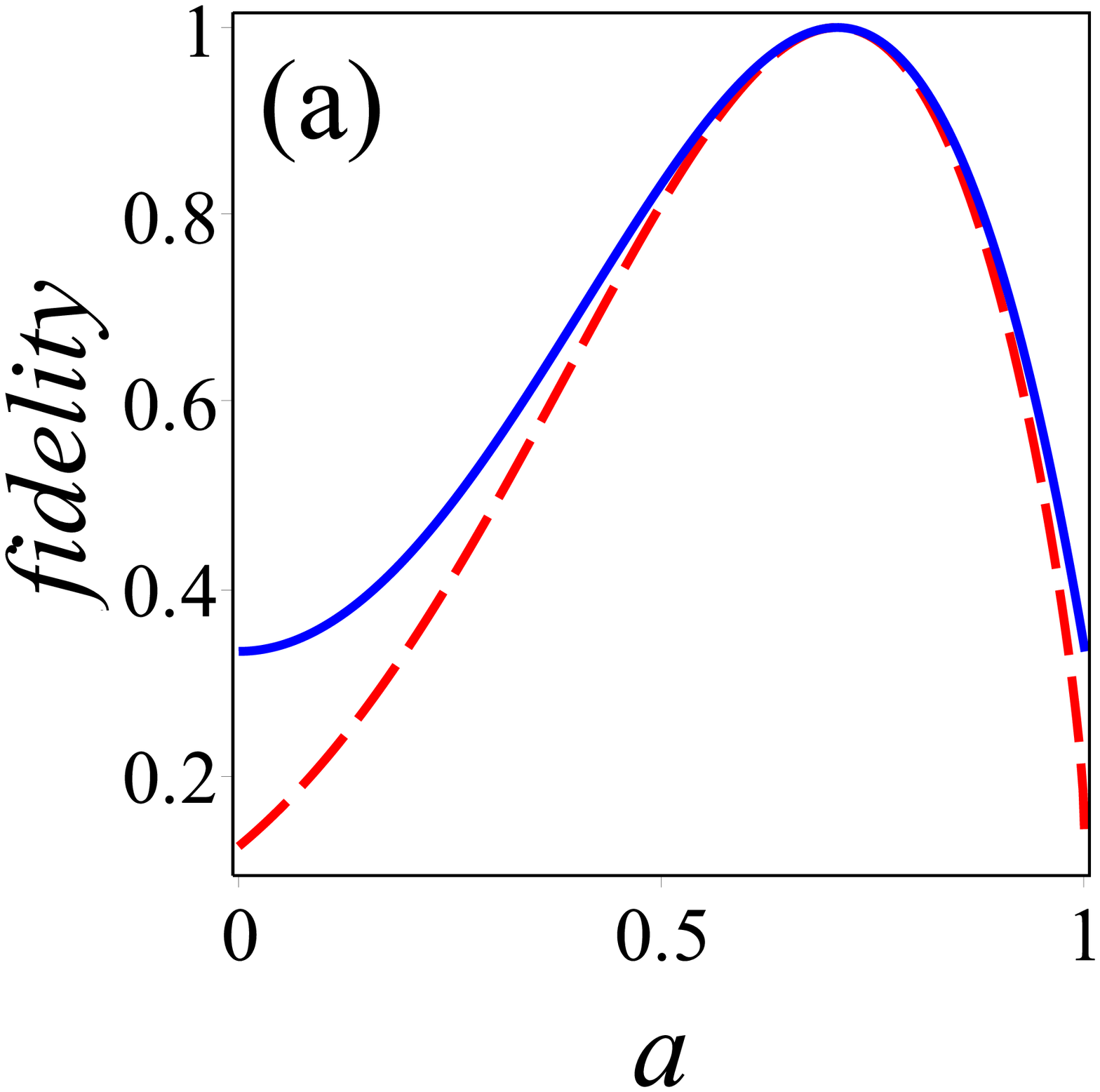}
\includegraphics[width=4.25cm]{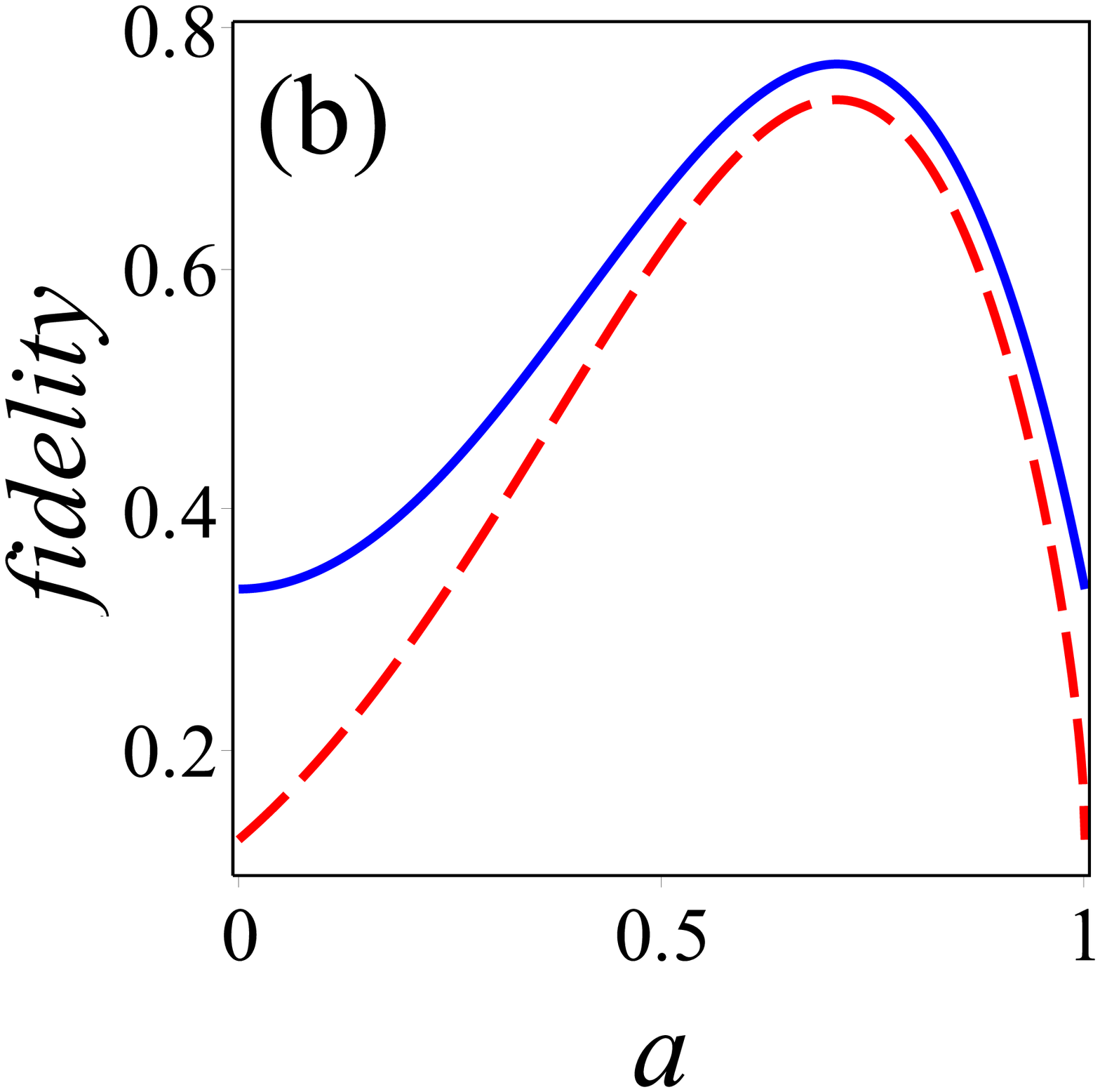}
\caption{(color online) Fidelities $F$ (solid) and $F_0$ (dashed) as functions of $a$ for (a) $\theta=0$  and (b) $\theta=\pi/5$. }
\label{Fig2}
\end{figure}

We proceed to show that even if all measurement outcomes are utilized, the fidelity of the resulting state with respect to $|W_1\rangle$ is slightly improved in comparison to the fidelity of (\ref{psi1}) with respect to the ideal source state $|\phi^+\rangle^{\otimes 3}$.
Without postselection, the possible states left in $B$ after the measurement in $A$ are listed in the first column of Table \ref{table2}.
The next step is to proceed with the operations in Table \ref{table1}. That leaves the first state in Table \ref{table2} unchanged, while the second (not normalized) state undergoes the transformation 
\begin{eqnarray}
\nonumber
(a^2+2b^2)|W_2\rangle+(a^2-b^2)(|W_3\rangle+|W_4\rangle) \longrightarrow\\
\nonumber
(a^2+2b^2)|W_1\rangle+(a^2-b^2)(|W_8\rangle-|W_7\rangle)\;,
\end{eqnarray}
and so on, see the second column of Table \ref{table2}. Gathering these results together and considering their probabilities (third column of Table \ref{table2}), the resulting density matrix for the mixed state reads, in the $W$ basis [Eq. (\ref{Wbasis})],
\begin{equation}
\rho=
\left( \begin{array}{cccccccc}
F & 0 & 0 & 0 & 0 & 0 & 0 & 0 \\
0 & 0 & 0 & 0 & 0 & 0 & 0 & 0 \\
0 & 0 & 0 & 0 & 0 & 0 & 0 & 0 \\
0 & 0 & 0 & 0 & 0 & 0 & 0 & 0 \\
0 & 0 & 0 & 0 & 0 & 0 & 0 & 0 \\
0 & 0 & 0 & 0 & 0 & \frac{1-F}{3} & -K & -K \\
0 & 0 & 0 & 0 & 0 & -K & \frac{1-F}{3} & -K \\
0 & 0 & 0 & 0 & 0 & -K & -K & \frac{1-F}{3} \\
\end{array} \right)\;,
\label{rho}
\end{equation}
where the absolute value of the coherences reads $K=[4a^2(1-a^2)(1+\sin^2\theta)+1]/9$ and
\begin{equation}
\label{fidW}
F=\frac{1}{3}(8\Gamma^2+1)\;,
\end{equation}
with $\Gamma=a\sqrt{1-a^2}\,\cos\theta$ and $b=\sqrt{1-a^2}e^{i \theta}$. The quantity in Eq. (\ref{fidW}) is the final fidelity, $F={\rm Tr}(\rho |W_1\rangle \langle W_1|)$. Since the described protocol amounts to lifting entanglement from a Hilbert space with dimension 4 to a Hilbert space with dimension 8, it could be considered useful even if the final fidelity were smaller than the fidelity of the source, to an acceptable extent. However, we note that the fidelity of $|\Phi_0\rangle$, Eq. (\ref{psi1}), with respect to an ideal source is $F_0=|\langle \Phi_0|(|\phi^+\rangle)^{\otimes 3}|^2$, or $F_0=\left(\Gamma+1/2\right)^3$,
with $-\pi/2 \le \theta \le \pi/2$. For phase arguments outside this range the source would better described by the Bell state $|\phi^-\rangle\sim |00\rangle - |11\rangle$. This leads to
\begin{equation}
F=\frac{2}{3}\left( 1-2F_0^{1/3}\right)^2+\frac{1}{3}\ge F_0\;.
\end{equation}
So that if the source produces single pairs with states $|00\rangle+e^{i\pi/5}|11\rangle$, whose fidelity is $F_0=(0.905)^3=0.74$, then we obtain an ensemble whose fidelity with respect to an ideal $W$ state is $F=0.77$ (see Fig. \ref{Fig2}).
In particular, a borderline initial fidelity of $0.5$ leads to $F=0.56$. 
It is important to recall that in this whole process, without postselection, there are no losses but those due to the measurement in $A$. Whenever the EPR source has $N$ particles the resulting $W$-state ensemble has $N/2$ particles, which renders statistical efficiency to the protocol. In addition, if the physical realization of the bipartite source is provided by entangled photons, we only demand first order PDC, that is, three pairs generated in sequence, provided that the setup is capable of keeping the coherence of the pairs in the typical window of time  for the occurrence of three PDC events.
Let us finally comment on the scenario in which the source, in addition to the systematic error, presents a fraction $f$ of white noise. A conservative hypothesis for the state of the triples in the source is
$\varrho_0=(1-f)|\Phi_0\rangle \langle \Phi_0| + f \,\mathbb{I}_{64}/64$, where $\mathbb{I}_D$ is the $D\times D$ identity operator. In this case it is immediate to see that the noise fraction remains unchanged throughout the protocol, and the obtained density operator reads $\varrho=(1-f)\rho + f\,\mathbb{I}_8/8$, where $\rho$ is given by Eq. (\ref{rho}). That is to say, our lifting procedure leads to entanglement concentration but probably not to entanglement distillation upon depolarizing noise. 
\section{Discussion and conclusion}
 It is arguably easier, both conceptually and experimentally, to create a four-partite $W$ state than to produce a tripartite $|W\rangle$. This is partly due to the fact that most of the measurements involved in the previous protocols are of Bell type. In this work we provide a full basis of $W$ states which induces more appropriate measurements to produce the desired type of entanglement. Of course, we are aware of the technical difficulties associated with full measurements involving entangled bases, either without interactions between the subsystems \cite{vaidman} or with passive linear elements only \cite{lut}. In the last fifteen years, many ways to circumvent these difficulties have been proposed, employing  ancillary systems (hyperentanglement) \cite{schuck,chieh}, nonlinear optics \cite{yoon}, or even active linear optical elements \cite{hussain}. In addition, we note that in the case of a perfect source, if one can distinguish $n$ basis elements out of 8, then we have a statistical efficiency of $n/8$. Even in the case of imperfect sources, to produce ideal $W$ states only requires the ability to distinguish $|W_1\rangle$ from $|W_5\rangle$, and these states from the rest of the basis elements. This is a much more modest task. Finally, we note that the measurement on subsystem $A$ can be made in a very localized region of space (without affecting the remote character of the state left in $B$), enabling the execution of CNOT gates involving particles 1, 3, and 5.

Our results can be summarized as follows: When the EPR source is ideal, a pure ensemble of $W$ states can be deterministically produced. In addition, if the bipartite source is non-maximally entangled, one can still obtain a set of pure $W$ states, this time at the cost of postselection. This is a peculiarity of the states $| W \rangle$ that, for instance, does not hold for GHZ states. In fact one can produce perfect $W$ states from arbitrarily poor entanglement, but, the poorest the entanglement of the source the smaller the number of states $|W\rangle $ in the ensemble.
Alternatively, it is possible to profit from every individual run, for, even without post selection one gets a mixed tripartite state whose fidelity (with respect to $|W\rangle$) is higher than that of the triple of pairs coming from the source (in comparison to a perfect triple of EPR pairs).
Finally we emphasize the possibility of creating remote entangled systems, allowing, e. g., distribution of keys.
\begin{acknowledgements}
Financial support from Conselho Nacional de Desenvolvimento Cient\'{\i}fico e Tecnol\'ogico (CNPq) through its program INCT-IQ, Coordena\c{c}\~ao de Aperfei\c{c}oamento de Pessoal de N\'{\i}vel Superior (CAPES), and Funda\c{c}\~ao de Amparo \`a Ci\^encia e Tecnologia do Estado de Pernambuco (FACEPE) is acknowledged.
\end{acknowledgements}


\begin{thebibliography}{90}




\bibitem{nilsen} M. Nielsen and I. Chuang, \textit{Quantum Computation and Quantum Information}, Cambridge, University Press (2010).


\bibitem{teleport} C. H. Bennet et al.:Teleporting an unknown quantum state via dual classical and Einstein-Podolsky-Rosen channels. Phys. Rev. Lett. {\bf 70}, 1895-1899 (1993).


\bibitem{nonlocal} N. Brunner, D. Cavalcanti, S. Pironio, V. Scarani, and S. Wehner.:Bell nonlocality. Rev. Mod. Phys. {\bf 86}, 419-478 (2014).


\bibitem{Raussendorf} R. Raussendorf and H. J. Briegel.:A One-Way Quantum Computer. Phys. Rev. Lett. {\bf 86}, 5188-5191 (2001).


\bibitem{ghz} D. M. Greenberger, M. A. Horne, A. Shimony, and A. Zeilinger.:Bell's theorem without inequalities. Am. J. Phys. {\bf 58}, 1131-1143 (1990).


\bibitem{ghz3} D. Bouwmeester, J.-W. Pan, M. Daniell, H. Weinfurter, and A. Zeilinger.:Observation of Three-Photon Greenberger-Horne-Zeilinger Entanglement. Phys. Rev. Lett. {\bf 82}, 1345-1349 (1999);

J.-W. Pan et al.:Experimental test of quantum nonlocality in three-photon Greenberger–Horne–Zeilinger entanglement. Nature {\bf 403}, 515 (2000); A. Rauschenbeutel et al.:Step-by-Step Engineered Multiparticle Entanglement. Science {\bf 288}, 2024-2028 (2000).


\bibitem{ghz4} C. A. Sackett et al.:Experimental entanglement of four particles. Nature {\bf 404}, 256-259  (2000).


\bibitem{ghz5} Z. Zhao et al.:Experimental demonstration of five-photon entanglement and open-destination teleportation. Nature {\bf 430}, 54-58 (2004).


\bibitem{ghz6} C.-Y. Lu et al.:Experimental entanglement of six photons in graph states. Nature Phys. {\bf 3}, 91-95 (2007).


\bibitem{ghz8} X.-C. Yao et al.:Observation of eight-photon entanglement. Nature Photon. {\bf 6}, 225-228 (2012).


\bibitem{dicke6} W. Wieczorek et al.:Experimental Entanglement of a Six-Photon Symmetric Dicke State. Phys. Rev. Lett. {\bf 103}, 020504 (2009);

R. Prevedel et al.:Experimental Realization of Dicke States of up to Six Qubits for Multiparty Quantum Networking. Ibid {\bf 103}, 020503 (2009).


\bibitem{xin} X.-W. Wang and G.-J. Yang.:Generation and discrimination of a type of four-partite entangled state. Phys. Rev. A {\bf 78}, 024301 (2008);

D. Das, S. Dogra, K. Dorai, and Arvind.:Experimental construction of a W superposition state and its equivalence to the Greenberger-Horne-Zeilinger state under local filtration. Ibid {\bf 92}, 022307 (2015).


\bibitem{Manfred} M. Eibl, N. Kiesel, M. Bourennane, C. Kurtsiefer, and H. Weinfurter.:Experimental Realization of a Three-Qubit Entangled W State. Phys. Rev. Lett. {\bf 92}, 077901 (2004).


\bibitem{mikami2} H. Mikami, Y. Li, K. Fukuoka, and T. Kobayashi.:New High-Efficiency Source of a Three-Photon W State and its Full Characterization Using Quantum State Tomography. Phys. Rev. Lett. {\bf 95}, 150404 (2005).


\bibitem{mikami} H. Mikami, Y. Li, and T. Kobayashi.:Generation of the four-photon W state and other multiphoton entangled states using parametric down-conversion. Phys. Rev. A {\bf 70}, 052308 (2004).


\bibitem{wieczorek} W. Wieczorek, N. Kiesel, C. Schmid, and H. Weinfurter.:Multiqubit entanglement engineering via projective measurements. Phys. Rev. A {\bf 79}, 022311 (2009).


\bibitem{tashima} T. Tashima et al.:Local Transformation of Two Einstein-Podolsky-Rosen Photon Pairs into a Three-Photon W State. Phys. Rev. Lett. {\bf 102}, 130502 (2009).


\bibitem{nmr} S. Dogra, K. Dorai, and Arvind.:Experimental construction of generic three-qubit states and their reconstruction from two-party reduced states on an NMR quantum information processor. Phys. Rev. A {\bf 91}, 022312 (2015).


\bibitem{Yamamoto} T. Yamamoto, K. Tamaki, M. Koashi, and N. Imoto.:Polarization-entangled W state using parametric down-conversion. Phys. Rev. A {\bf 66}, 064301 (2002).


\bibitem{cheng} Y.-B. Sheng, F.-G. Deng, and H.-Y. Zhou.:Generation of Multiphoton Entangled States with Linear Optical Elements. Chin. Phys. Lett. {\bf 25}, 3558 (2008).


\bibitem{tashima1} T. Tashima et al.:Elementary optical gate for expanding an entanglement web. Phys. Rev. A {\bf 77}, 030302 (2008).


\bibitem{tashima2} T. Tashima et al.:Local expansion of photonic W state using a polarization-dependent beamsplitter. New J. Phys. {\bf 11}, 023024 (2009).


\bibitem{tashima3} T. Tashima et al.:Demonstration of Local Expansion Toward Large-Scale Entangled Webs. Phys. Rev. Lett. {\bf 105}, 210503 (2010).


\bibitem{xu} W. Xu, X. Zhao, and G. Long.:Efficient generation of multi-photon W states by joint-measurement. Prog. Nat. Sci. {\bf 18}, 119-122 (2008).


\bibitem{yang} R.-C. Yang and T.-C. Zhang.:Robust preparation of atomic W states without any excitations. Opt. Comm. {\bf 284}, 3164-3167 (2011).


\bibitem{tong} X. Tong et al.:A hybrid-system approach for W state and cluster state generation. Opt. Comm. {\bf 310}, 166-172 (2014).


\bibitem{ozaydin} C. Yesilyurt et al.:An Optical Setup for Deterministic Creation of Four Partite W state. Acta Phys. Pol. A {\bf 127}, 1230 (2015).


\bibitem{yesil} C. Yesilyurt et al.:Deterministic Local Expansion of W States. arXiv: 1602.04166 (2016).


\bibitem{fusion0} S. K. \"Ozdemir et al.:An optical fusion gate for W-states. New J. Phys. {\bf 13}, 103003 (2011);

C. Yesilyurt, S. Bugu, and F. Ozaydin.:An optical gate for simultaneous fusion of four photonic
W or Bell states. Quantum Inf. Proc. {\bf 12}, 2965-2975 (2013);

S. Bugu, C. Yesilyurt, and F. Ozaydin.:Enhancing the W-state quantum-network-fusion process with a single Fredkin gate. Phys. Rev. A {\bf 87}, 032331 (2013);

F. Ozaydin et al.:Fusing multiple W states simultaneously with a Fredkin gate. ibid {\bf 89}, 042311 (2014);

X.-P. Zang et al.:Generating multi-atom entangled W states via light-matter interface based fusion mechanism. Scientific Reports {\bf 5}, 16245 (2015).

K. Li, M. Yang, Q. Yang, and Z.-L. Cao.:Fusion of W-like states in optical system. Laser Phys. {\bf 26}, 025203 (2016).


\bibitem{feihu} F. Xu et al.:Experimental quantum key distribution with source flaws. Phys. Rev. A {\bf 92}, 032305 (2015).


\bibitem{mizutani} A. Mizutani, N. Imoto, and K. Tamaki.:Robustness of the round-robin differential-phase-shift quantum-key-distribution protocol against source flaws. Phys. Rev. A {\bf 92}, 060303(R) (2015).




\bibitem{pulse-mode} S. K. \"Ozdemir, A. Miranowicz, M. Koashi, and N. Imoto.:Pulse-mode quantum projection synthesis: Effects of mode mismatch on optical state truncation and preparation. Phys. Rev. A {\bf 66}, 053809 (2002).


\bibitem{scherer} A. Scherer, R. B. Howard, B. C. Sanders, and W. Tittel.:Quantum states prepared by realistic entanglement swapping. Phys. Rev. A {\bf 80}, 062310 (2009).


\bibitem{wei} J-W Pan and A. Zeilinger.:Greenberger-Horne-Zeilinger-state analyzer. Phys. Rev. A {\bf 57},2208 (1998).


\bibitem{conc} C. Bennett, H. J. Bernstein, S. Popescu, and B. Schumacher.:Concentrating partial entanglement by local operations. Phys. Rev. A {\bf 53}, 2046 (1996).


\bibitem{bose} S. Bose, V. Vedral, and P. L. Knight.:Purification via entanglement swapping and conserved entanglement. Phys. Rev. A {\bf 60}, 194 (1999).


\bibitem{dur} W. D\"ur, G. Vidal, and J. I. Cirac.:Three qubits can be entangled in two inequivalent ways. Phys. Rev. A {\bf 62}, 062314 (2000).


\bibitem{vaidman} L. Vaidman and N. Yoran.:Methods for reliable teleportation. Phys. Rev. A {\bf 59}, 116 (1999).


\bibitem{lut} N. L\"utkenhaus, J. Calsamiglia, and K.- A. Suominen.:Bell measurements for teleportation. Phys. Rev. A {\bf 59}, 3295 (1999).


\bibitem{schuck} C. Schuck, G. Huber, C. Kurtsiefer, and H. Weinfurter.:Complete Deterministic Linear Optics Bell State Analysis. Phys. Rev. Lett. {\bf 96}, 190501 (2006).


\bibitem{chieh} T.-C. Wei, J. T. Barreiro, and P. G. Kwiat.:Hyperentangled Bell-state analysis Phys. Rev. A {\bf 75}, 060305(R) (2007);

Y.-B. Sheng and F.-G. Deng.:Deterministic entanglement purification and complete nonlocal Bell-state analysis with hyperentanglement. ibid {\bf 81}, 032307 (2010);

N. Pisenti, C. P. E. Gaebler, and T. W. Lynn.:Distinguishability of hyperentangled Bell states by linear evolution and local projective measurement. ibid {\bf 84}, 022340 (2011);

W. P. Grice.:Arbitrarily complete Bell-state measurement using only linear optical elements ibid {\bf 84}, 042331 (2011).


\bibitem{yoon} Y.-H. Kim, S. P. Kulik, and Y. Shih.:Quantum Teleportation of a Polarization State with a Complete Bell State Measurement. Phys. Rev. Lett. {\bf 86}, 1370 (2001).


\bibitem{hussain} H. A. Zaidi and P. van Loock.:Beating the One-Half Limit of Ancilla-Free Linear Optics Bell Measurements. Phys. Rev. Lett. {\bf 110}, 260501 (2013).


\end{thebibliography}
\end{document}